\documentclass[preprint,showpacs,preprintnumbers,amsmath,amssymb]{revtex4}
\usepackage{blindtext}
\usepackage{graphicx}
\usepackage{amssymb}
\usepackage{amsfonts}
\usepackage{amsmath}
\usepackage{textcomp}
\usepackage{xcolor}
\usepackage{ulem}
\usepackage{bm}
\usepackage{parskip,float,graphicx}
\usepackage[caption = false]{subfig}
\usepackage[colorlinks=true,allcolors=blue]{hyperref}
\usepackage{physics}
\usepackage[separate-uncertainty=true,exponent-product=\!\cdot\!]{siunitx}
\newcommand{\e}{\mathrm{e}}

\newcommand{\ntot}{n_\text{tot}}
\newcommand{\Jxst}{J_x^\text{st}}
\newcommand{\Jyst}{J_y^\text{st}}
\newcommand{\nst}{n^\text{st}}
\newcommand{\Jxsurf}{J_x^\mathsf{S}}
\newcommand{\funcd}{\boldsymbol{\delta}}
\begin{document}

\title{Surface currents in Hall devices}
\author{M. Creff$^{1,2}$}  \author{F. Faisant$^{1}$}  \author{J. M. Rub\`i$^{2}$} \author{J.-E. Wegrowe$^1$}
\affiliation{$^1$ LSI, \'Ecole Polytechnique, CEA/DRF/IRAMIS, CNRS, Institut Polytechnique de Paris, F-91128 Palaiseau, France}
\affiliation{$^2$ Department de Fisica Fonamental, Universitat de Barcelona, Spain}
\date{\today}


\date{\today}

\begin{abstract}

A variational approach is used in order to study the stationary states of Hall devices. Charge accumulation, electric potentials and electric currents are investigated on the basis of the Kirchhoff-Helmholtz principle of least heat dissipation. A simple expression for the state of minimum power dissipated -- that corresponds to zero transverse current and harmonic chemical potential -- is derived. It is shown that a longitudinal surface current proportional to the charge accumulation is flowing near the edges of the device. Charge accumulation and surface currents define a boundary layer over a distance of the order of the Debye-Fermi length. 

\end{abstract}


\maketitle

The description of the classical Hall effect \cite{Hall} (i.e.\ in the diffusive limit) is usually based on the local transport equations for the charge carriers under both an electric field and the Laplace-Lorentz force generated by a static magnetic field  $\vec{H}$. The physical mechanisms behind this effect and the corresponding transport equations are well-known and are described in all reference textbooks \cite{Kittel,McGraw, Sze, Aschcroft, Popovic}. The stationarity condition is set independently through the local continuity equation, namely by imposing a divergenceless electric current: $\vec \nabla \cdot \vec J^\text{st} = 0$. However, this stationarity condition is not sufficient to describe the accumulation of electric charges at the edges, and the determination of the boundary conditions of the Hall effect is still an open problem \cite{boundary,Calcul,Nanowire,Benda}. In particular, the electric charges accumulated at the edges are not static and the existence of surface currents -- which is intuitively expected -- has been overlooked and does not seem to have been the object of devoted works. After one hundred forty years of intensive studies and technological developments of Hall devices, we suspect that this surprising situation is due to the limitations of the local stationarity condition mentioned above.

The goal of the present work is to reconsider the stationary states of the Hall effect in a variational framework -- namely the Kirchhoff-Helmholtz principle of least heat dissipation \cite{Benda, EPL1,EPL2} -- in order to characterize the system globally, including the edges and beyond.

The use of a global -- instead of a local -- stationarity condition is not without practical consequences as it allows to explain why a stable Hall voltage can be measured in conventional Hall devices -- despite lateral leak of electric charges due e.g. to the presence of a voltmeter -- by renewing permanently the electric charges accumulated at the edges. 

The system is first defined from a thermodynamic point of view, and  the minimization of the Joule power under both electrostatic screening and galvanostatic constraints is then performed. A differential equation for the current density is obtained. The state of least dissipation is then derived and discussed.\\

\begin{figure} [h!]
   \begin{center} 
   \begin{tabular}{c}
\includegraphics[height=6cm]{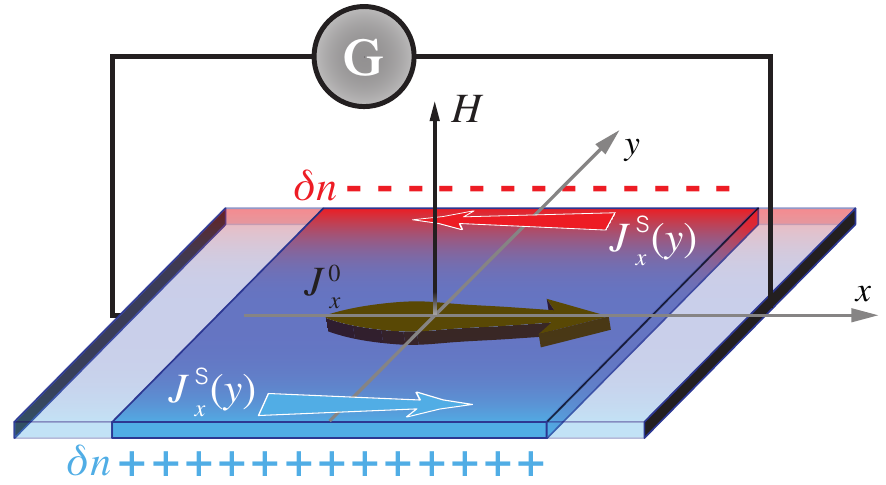}
   \end{tabular}
   \end{center}
\caption[Fig1]
{ \label{fig:Fig1} Schematic representation of the Hall effect under a static magnetic field $H$ applied along the $z$ direction, with the electrostatic charge accumulation $\delta n$ and surface currents $\Jxsurf(y)=J_x-J_x^0$ at the edges.}
\end{figure}

The system under interest is defined in the context of non-equilibrium thermodynamics \cite{DeGroot,Mazur,Rubi}. It is a thin conducting layer of finite width forced to an electric generator and submitted to a magnetic field (see Fig.\,\ref{fig:Fig1}). We assume the invariance along the $x$ axis (in particular, the region in contact to the generator is not under consideration here). 
However, it is important to point-out that electric charges are allowed to flow from one lateral edge to the other in order to take into account the Hall voltage measurements  (since any realistic voltmeter has a finite internal resistance).

Let us define the distribution of electric charge carriers by $n(y) = n_0 + \delta n(y)$, where $\delta n(y)$ is the charge accumulation and $n_0$ the homogeneous density of an electrically neutral system (e.g. density of carriers without the magnetic field). The charge accumulation is governed by the Poisson's equation $\nabla^2 V = \partial_y^2 V = -\frac{q}{\epsilon} \delta n$, where $V$ is the electrostatic potential, $q$ is the electric charge, and $\epsilon$ is the electric permittivity. The local electrochemical potential $\mu(x,y)$, that takes into account not only the electrostatic potential $V$ but also the energy due to the charge accumulation $\delta n$, is given by \cite{Mazur,Rubi} (local equilibrium is assumed everywhere) : 
\begin{equation}
  \mu = \frac{k T}{q} \ln \left( \frac{n}{n_0} \right) + V
  \label{mu}
\end{equation}
where $k$ is the Boltzmann constant and the temperature $T$ is the Fermi temperature $T_F$ in the case of a fully degenerated conductor, or the temperature of the heat bath in the case of a non-degenerated semiconductor \cite{Rque}. Poisson's equation now reads
\begin{equation}
  \nabla^2 \mu - \lambda_D^2 \frac{q}{\epsilon} n_0 \nabla^2 \ln \left( \frac{n}{n_0} \right) + \frac{q}{\epsilon} \delta n = 0
  \label{poisson-mu}
\end{equation}
where $\lambda_D= \sqrt{\frac{k T \epsilon}{q^2 n_0}}$ is the Debye-Fermi length. The invariance along $x$ gives $\nabla^2 = \partial_y^2$.

On the other hand, the transport equation under a magnetic field is $\vec{J} = -\hat{\sigma} \vec{\nabla} \mu = - q n \hat{\eta} \vec{\nabla} \mu$, with the conductivity tensor $\hat{\sigma}$ and the mobility tensor $\hat{\eta}$. In two dimensions and for isotropic material, the mobility tensor is defined by Onsager relations \cite{Onsager}:
\begin{equation*}
  \hat{\eta} =
  \begin{bmatrix}
    \eta    & \eta_H \\
    -\eta_H & \eta
  \end{bmatrix}
  = \eta
  \begin{bmatrix}
    1         & \theta_H \\
    -\theta_H & 1
  \end{bmatrix}
  \quad \text{ with } \quad \theta_H = \frac{\eta_H}{\eta}
\end{equation*}
where $\eta$ is the ohmic mobility, $\eta_H$ the Hall mobility (function of the magnetic field $\vec{H}=H\vec{e}_z$) and $\theta_H$ the Hall angle. 
The electric current then reads: $ \vec{J} = - q n \eta \left( \vec{\nabla} \mu - \theta_H \, \vec{e_z} \wedge \vec{\nabla} \mu \right)$, or:
\begin{align}
  -q n \eta (1+\theta_H^2) \partial_x \mu & = J_x - \theta_H J_y 
  \label{relations-dxmu} \\
  -q n \eta (1+\theta_H^2) \partial_y \mu & = J_y + \theta_H J_x 
  \label{relations-dymu}
\end{align}
The Kirchhoff-Helmholtz principle states that the current distributes itself so as to minimise Joule heating $\vec J \cdot \vec \nabla \mu$ on a domain $\mathcal D$:
\begin{equation*}
  P_J  = \int_{\mathcal D}  q n \eta \|\vec{\nabla} \mu\|^2 \dd{x}\dd{y}
       = \frac{1}{q n_0 \eta (1+\theta_H^2)} \int_{\mathcal D} \frac{n_0}{n} \|\vec{J}\|^2 \dd{x}\dd{y}
\end{equation*}

After introducing the galvanostatic contraint $J_x = -q n (\eta \partial_x \mu + \eta_H \partial_y \mu)$, the screening equation (\ref{poisson-mu}) and their respective Lagrange multipliers $\beta$ and $\gamma$, the functional to be minimized reads:
\begin{equation}
\begin{aligned}
F[n,\vec{\nabla} \mu] = \int_{\mathcal D} q n \eta \|\vec{\nabla} \mu\|^2 \dd{y} - \int_{\mathcal D} \beta(y) \left ( - q \eta n\partial_x \mu - q \eta_H n \partial_y \mu  \right ) \dd{y} \\ - \int_{\mathcal D} \gamma(y) \left ( \nabla^2 \mu - \lambda_D^2 \frac{q}{\epsilon} n_0 \nabla^2 \ln \left( \frac{n}{n_0} \right) + \frac{q}{\epsilon} \delta n \right ) \dd{y}
\end{aligned}
\label{Functional}
\end{equation}
The minimization imposes the vanishing of the functional derivatives $ \frac{\funcd F}{\funcd (\partial_x \mu)} $,  $\frac{\funcd F}{\funcd (\partial_y \mu)} $ and $\frac{\funcd F}{\funcd (n)} $, from which we obtain the Euler-Lagrange equation corresponding to the stationary state (see supplemental material \cite{SupMat}):
\begin{equation}
J_y - \lambda_D^2 \partial_y \left( \frac{n_0}{n} \partial_y J_y \right) = \frac{\varepsilon}{2 q^2 \eta (1+\theta_H^2)} \partial_y \left( \frac{(J_y)^2 + 2 \theta_H J_x J_y - (J_x)^2}{n^2} \right)
\label{Eq1}
\end{equation}
This is a second order differential equation in $J_y$ and first order in $n$, and its resolution -- coupled with Poisson's equation and transport equations-- would need the knowledge of four boundary conditions. However, these conditions are not imposed externally but are fixed by the system itself in order to reach the state of minimum dissipation.
Our approach does not consider them explicitly, because the functional (\ref{Functional}) should also include the treatment of the discontinuity between the conductor its environment. 

Fortunately, it is possible to bypass this difficulty by finding the minimum power dissipated without the aforementioned boundary conditions while taking into account the global conserved quantities, which are known. Let us define the width $L$ of the conductor and the two following global quantities: the total charge carrier density $\ntot = \frac{1}{L}\int n \dd{y} $ (we expect $\ntot = n_0$ for global charge neutrality), the global current flowing in the $x$ direction throughout the device $J_x^0 = \frac{1}{L} \int J_x \dd{y}$ (which is constant along $x$ by definition of the galvanostatic condition). Furthermore, let us define for convenience $\delta J_x = J_x - J_x^0 \frac{n}{\ntot}$ (which is globally null in the sense that $\int J_x \dd{y} - \frac{J_x^0}{\ntot} \int n \dd{y} = 0$) and $\tilde{P_J} = P_J q \eta (1+\theta_H^2) = \int_D \frac{\|\vec{J}\|^2}{n} \dd{y}$. We then have:
\begin{equation*}
  \Tilde{P_J} = \int \frac{J_x^2 + J_y^2}{n} \dd{y}  = \frac{(J_x^0)^2}{\ntot}L + \underbrace{ 2 \frac{J_x^0}{\ntot} \int \delta J_x \dd{y} }_{=0} + \underbrace{ \int \frac{(\delta J_x)^2 + J_y^2}{n} \dd{y} }_{\geqslant 0},
\end{equation*}
so that the dissipation is always such that $ \Tilde{P_J}(n, J_x, J_y) \geqslant \frac{(J_x^0)^2}{\ntot}L$, i.e. greater than in the situation for which $J_y=0$. Hence, the minimum is reached for 
\begin{equation}
  \Jxst(y) = J_x^0\frac{n(y)}{\ntot} \quad \text{and} \quad \Jyst = 0
  \label{min}
\end{equation}
It is easy to verify that this state is a solution of the Euler-Lagrange equation (\ref{Eq1}), whatever the density distribution $n(y)$. Furthermore, as shown below, this solution is stable. Note also that the usual stationarity condition $\vec \nabla \cdot \vec J^\text{st} = 0$ is verified, but now we have $\vec \nabla \times \vec J^\text{st} \neq 0$. Inserting the solution (\ref{min}) into the relations (\ref{relations-dxmu},\ref{relations-dymu}), we deduce $\partial_x \mu^\text{st} = \frac{-J_x^0}{q \ntot \eta (1+\theta_H^2)}$ and $\partial_y \mu^\text{st} = \frac{\theta_H J_x^0}{q \ntot \eta (1+\theta_H^2)}$. These two terms are constant so that the electochemical potential of the stationary state is harmonic: $\nabla^2 \mu^\text{st} = 0$. The corresponding current (\ref{min}) is defined as a function of the charge density $n$. The solution is hence given by Poisson's equation for $\nabla^2 \mu^\text{st} = 0$:
\begin{equation}
  \lambda_D^2 \partial_y^2 \ln \left(1 + \frac{\delta \nst }{n_0} \right) = \frac{\delta \nst}{n_0}
  \label{poisson-final}
\end{equation}
It is still necessary to know two boundary conditions on $n$ in order to determine the solution. Once again these boundary conditions are not explicitely given, but we can use global conditions instead. The first one is given by $\int n \dd{y} = \ntot L$, and a second condition is imposed by the expression of the electric field $E_y$, given by Gauss's law at a point $y_0$ (see Supplemental material \cite{SupMat}):
\begin{equation*}
  E_y(y_0) = - \partial_y V(y_0) = - \frac{q}{2 \varepsilon} \int_{-L/2}^{L/2} \delta n(y) \frac{y-y_0}{\abs{y-y_0}} \dd{y} + E(\infty) + E(- \infty)
\end{equation*}
whose derivative is nothing but the Poisson's equation. The constant $E(\infty) + E(- \infty)$ accounts for the electromagnetic environment of the Hall device ($E(\pm\infty) = 0$ in vacuum). Inserting (\ref{min}) and the relation (\ref{relations-dymu}) for $\partial_y\mu$ gives the final condition:
\begin{equation}
  \frac{2 \theta_H J_x^0 C_0}{1 + \theta_H^2} + 2 \lambda_D^2 \partial_y \ln(\frac{\nst}{n_0})(y_0) +2 C_E +  \int_{-L/2}^{L/2} \delta \nst(y) \frac{y-y_0}{\abs{y-y_0}} \dd{y} = 0
  \label{condition}
\end{equation}
where $C_0 = \frac{\varepsilon }{q^2 \ntot \eta}$ and $C_E = \frac{\epsilon (E(\infty) + E(- \infty))}{q n_0}$. The sign of $(\delta \nst(y))(y-y_0)$ is fixed by the sign of $\theta_H J_x^0$ meaning that the side where $\delta \nst > 0$ is fixed by the direction of the current in $x$ and by the magnetic field. Using this condition and fixing $n_{tot}$ gives a unique solution for $\nst$ and the surface currents (\ref{min}) are now fully determined.

\begin{figure}[H]
    \centering
    \includegraphics[width=0.6\textwidth]{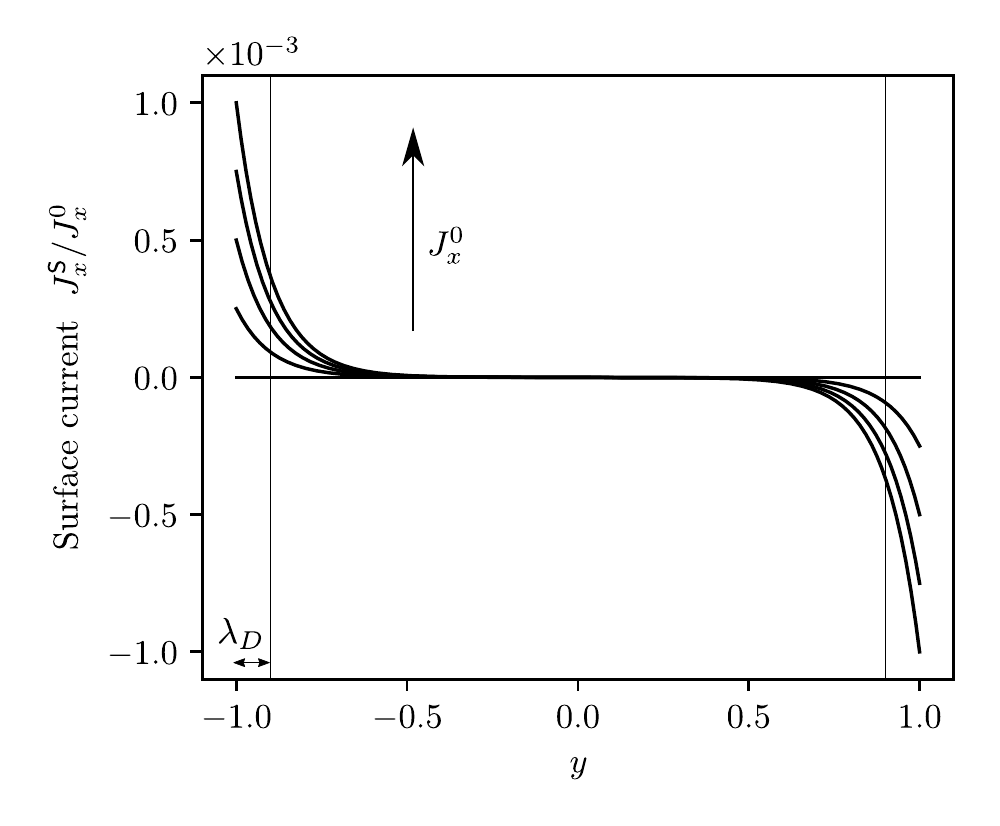}
    \caption{Numerical solutions for the surface currents  (equation (\ref{condition})).  The sample is confined in the region $y \in [-1,1]$. The straight vertical lines represent $\lambda_D$. The five profiles correspond to different values of the Hall angles $\theta_H$ and the parameter $C_0$ ($C_E = 0$).}
  \end{figure}

The linearization of Eqs. (\ref{poisson-final},\ref{condition}) when $\delta \nst \ll n_0$ gives an analytical solution of the problem. We then have $\frac{\delta \nst}{n_0} = A \e^{\frac{y}{\lambda_D}} + B \e^{-\frac{y}{\lambda_D}}$. The condition of neutrality ($\ntot = n_0 \Leftrightarrow \int \delta \nst \dd{y} = 0$) imposes that $A = -B$, and we can use the condition (\ref{condition}) to determine $A$ at $y_0 = 0$. 
Then, using the stationary solution $\Jxst \propto n$ (\ref{min}), the expression of $\delta \nst$ defines a {\it surface current} $\Jxsurf(y) = \Jxst - J_x^0$, which is superimposed to the galvanostatic current $J_x^0$: 
\begin{equation}
  \Jxsurf(y) = - J_x^0 \, \,\frac{1}{ \lambda_D} \left ( \frac{J_x^0 \, C_0 \theta_H}{n_0 (1+\theta_H^2)} + C_E \right) \, \frac{\sinh \big( \frac{y}{\lambda_D} \big)}{\cosh \big( \frac{L}{2 \lambda_D} \big) }.
  \label{surface-current}
\end{equation}
 Note that for $C_E = 0$, the surface currents are proportional to the square of the injected current ${J_x^0}$: this characteristic may be checked experimentally. 

For a vanishing screening length $\lambda_D \rightarrow 0$, we obtain the Dirac distribution
\begin{equation*}
  \Jxsurf(y) = J_x^0 \, \, \ \left ( \frac{J_x^0 \, C_0 \theta_H}{n_0 (1+\theta_H^2)} + C_E \right) \, \left( \delta_{-\frac{L}{2}}(y) - \delta_{+\frac{L}{2}}(y) \right)
\end{equation*}
When $\theta_H \ll 1$ and $E(\pm\infty) = 0$, the Dirac charge accumulation creates surface charge $Q_s = \frac{qC_0 J_x^0}{n_0^2} \theta_H$. The voltage then reads $V_H = \frac{Q_s L}{\epsilon} = \frac{\theta_H L J_x^0}{q n_0 \eta}$ and the usual formula for the Hall voltage is recovered by taking $\theta_H = \arctan(\eta H) \simeq \eta H$. \\

Finally, it is important to verify that the solution defined by Eq.\,(\ref{min}) is regular or stable enough, so that the system can relax from the transitory regime to this stationary state. 
According to the Kirchhoff-Helmholtz principle (i.e. the second law of thermodynamics), this is the case if the system can reach the minimum heat production, i.e. in the framework of our model, if a solution of Eq.\,(\ref{Eq1}) (and the Poisson's equation) can converge uniformly to the minimum defined by Eq.\,(\ref{min}). 
This is indeed the case:
with a given $J_y$, $J_x$ and $n$, let us define $\epsilon_y = J_y/n$ , $\epsilon_x = J_x/n - J_x^0/n_0$, $ n_+ = n - \nst $ (with $\nst$ the solution of Eq.\,(\ref{poisson-final})) and 
$f(y) = \frac{\delta n}{n_0}  - \lambda_D^2 \partial_y^2 \ln \Big(\frac{n}{n_0}\Big)$. It is clear that $f(y)$ tends to 0 when $n$ tends to $\nst$.  Furthermore, Poisson's equation now reads:
 \begin{equation}
 \partial_y \epsilon_y + \theta_H \partial_y \epsilon_x = f(y).
 \label{Poisson2}
 \end{equation}
This equation shows that if both $f(y)$ and $\epsilon_y$ tend to $0$  then $\epsilon_x$ tends to a constant. The galvanostatic constraint shows that this constant is $0$. In the same way, if both $\epsilon_x$ and  $\epsilon_y $ tend to  $0$, then $f(y)$ tends to $0$ and $n$ tends to $\nst$. 
Finally, Eq.\,(\ref{Eq1}) now reads:
\begin{equation}
\begin{aligned}
\left (n  - \frac{\varepsilon}{q^2 \eta} f(y) \right )\epsilon_y = \frac{\varepsilon}{2q^2 \eta}   \left ( (1 + \theta_H^2) \, \partial_y \left ( \frac{J_x^0}{n_0} + \epsilon_x \right )^2 + 2 \theta_H \frac{J_x^0}{n_0} \left ( f(y) - \theta_H \, \partial_y \epsilon_x  \right ) \right ) ,
\end{aligned}
\end{equation}
which shows that if both $f(y) $ and $\epsilon_x$ tend to $0$ then $\epsilon_y$ tends to $0$. \\

In conclusion, we have shown that the stationary state of the Hall effect is characterized not only by an accumulation of electric charges $Q(y) = q \delta n(y) = q (n(y) - n_0)$ at the edges - that generates the Hall voltage as in a simple capacitor at equilibrium - but also by surface currents that are flowing in opposite directions along the edges, and that are proportional to the charge accumulation: $J_x(y) = J_x^0 \,(1 + \delta n(y)/n_0)$ (see Fig.1). These currents describe the fact that the accumulation of electric charges $\delta n$ is not at equilibrium, but is renewed permanently by the generator. A simple expression of the surface current is given for $\delta n/n_0 \ll 1$ (\ref{surface-current}) and the usual Hall voltage is recovered for small magnetic fields. 
 
\section{ Aknowlegement}
 We thank Robert Benda, Jean-Michel Dejardin and Serge Boiziau for their important contributions to early developments of this work.


\begin{thebibliography}{0}
\bibitem{Rque} The expression is valid for non-degenerated semiconductor (Maxwellian distribution). However, in the case of degenerated metal, the expression is an approximation for $\delta n/n_0  \ll 1$
\end{thebibliography}
\end{document}